\documentclass[twocolumn,floats,pra,showpacs]{revtex4-1}

\usepackage{amsmath}
\usepackage{amsfonts}

\usepackage{tikz}

\newcommand {\be}{\begin{equation}}
\newcommand {\ee} {\end{equation}}
\newcommand {\bea}{\begin{eqnarray}}
\newcommand {\eea} {\end{eqnarray}}

\newcommand{\half}{{\textstyle{\frac{1}{2}}}}

\begin{document}


\title{Comment on ``Galilean invariance at quantum Hall edge''}
\author{J. H\"oller and N. Read}
\affiliation{Department of Physics, Yale
University, P.O. Box 208120, New Haven, CT 06520-8120, USA}
\date{May 1, 2016}

\begin{abstract}
In a recent paper by S. Moroz, C. Hoyos, and L. Radzihovsky [Phys. Rev. B {\bf 91}, 195409 (2015)], it
is claimed that the conductivity at low frequency $\omega$ and small wavevector $q$ along the edge of a
quantum Hall (QH) system (that possesses Galilean invariance along the edge) contains a universal 
contribution
of order $q^2$ that is determined by the orbital spin per particle in the bulk of the system, or
alternatively by the shift of the ground state. (These quantities are known to be related to the Hall
viscosity of the bulk.) In this Comment we calculate the real part of the
conductivity, integrated over $\omega$, in this regime for the edge of a system of non-interacting electrons
filling either the lowest, or the lowest $\nu$ ($\nu=1$, $2$, \ldots),  Landau level(s), and show that 
the $q^2$ term is non-universal and depends on details of the confining potential at the edge. In the 
special case of a linear potential, a form similar to the prediction is obtained; it is possible that 
this corrected form of the prediction may also hold for fractional QH states in systems with special forms 
of interactions between electrons.
\end{abstract}


\maketitle

\section{Introduction}

The Hall viscosity \cite{asz} in the bulk of a quantized Hall state of a two-dimensional electron gas is 
now known to be given by
\be
\eta^H=\half\bar{s}\bar{n}\hbar,
\label{hallviscrel}
\ee
when rotational invariance (as well as translation invariance) is present \cite{read09,rr11}. Here 
$\bar{n}$ is the particle density, and $\bar{s}$ is (minus) the ``mean orbital spin per particle''. 
The latter is also related \cite{read09,rr11} to the shift $\cal S$, which can be defined by
\be
N_\phi=\nu^{-1}N-{\cal S},
\ee
where $\nu$ is the filling factor of the quantum Hall state, $N$ is the particle number, and $N_\phi$ is
the number of flux; $N$ and $N_\phi$ are determined as values at which the ground state on the sphere is
free of defects (quasiparticle excitations). The relation is ${\cal S}=2\bar{s}$ \cite{wz92}. In addition, 
when the system is Galilean invariant, the density of kinetic momentum is proportional to the current 
density, and there is a relation of the $q^2$ part of the Hall conductivity with the Hall viscosity 
\cite{hoyos,bgr}.

In a recent paper \cite{mhr}, Moroz {\it et al.}\/ suggested that in a Galilean-invariant system, the 
orbital spin per particle, or shift, also appears as a universal coefficient in the $q^2$ term of the 
{\em longitudinal} conductivity of the system at an edge. 
Their prediction for the low-frequency, small wavevector region (with the temperature equal to zero) 
is
\be
\sigma_{yy}(q,\omega)=\frac{\nu}{2\pi}\left(1+\frac{\cal S}{4}\frac{q^2}{B}\right)\frac{-iv_F}
{\omega^+-qv_F} - im\epsilon''(B)\frac{q}{B}.
\label{mhr_pred1}
\ee 
The notation, with minor changes from Ref.\ \cite{mhr}, is as follows: $\sigma_{yy}$ is the longitudinal 
conductivity along the edge, which runs parallel to the $y$-axis; $q$ is the wavevector, parallel to the 
$y$-axis;
$\omega^+=\omega+i0^+$; $v_F>0$ is the velocity of edge excitations in the limit of low excitation energy;
$\epsilon''(B)$ is a real function of the magnetic field $B$, and $m$ is the particle mass. Throughout,
we will set $\hbar=1$, absorb the charge of the electron into the magnetic field, and take $B>0$. They
suggested that this relation might make it possible to measure $\cal S$.

Taking the real part using the Sokhotski-Plemelj relation, we obtain the claim that we wish to study,
\be
{\rm Re}\,\sigma_{yy}(q,\omega)= -\half \nu v_F\left(1+\frac{\cal 
S}{4}\frac{q^2}{B}\right)\delta(\omega-qv_F).
\label{mhr_pred2}
\ee
Some difficulties are already apparent with this formula. First, the real part of the conductivity (which 
we also call the spectral density) should be positive. If the sign is corrected, the order $q^0$ part is 
the expected universal form  \cite{wen}, related to the Hall conductivity of the bulk 
$\sigma_{xy}=\nu/(2\pi)$ in our units \cite{ch}. The second problem arises with the 
$q^2\delta(\omega-qv_F)$ term: in the absence of Lorentz invariance, edge excitations \cite{halperin} do 
not generally have a perfectly
linear energy versus momentum relation; instead there is usually some curvature. This does not affect 
the order $q^0$ part (we will assume there is a single velocity for asymptotically 
low-energy excitations), but at higher order in $q$ it usually leads to a form for the spectral density 
that is no longer a $\delta$-function of frequency, but instead is spread over a range of frequencies 
of order $q^2$ as $q\to0$, an effect omitted in this formula. As such an effect
definitely occurs in some of the models that we will study, it is not even clear exactly how to relate a
calculated spectral density to the predicted $\delta$-function. A natural (but as we
will discuss later, not unique) way to do so is to calculate the total spectral weight at each $q$. Thus,
we will calculate the integral of ${\rm Re}\,\sigma_{yy}$ over frequency (in the low-frequency region), and
compare the result with the coefficient of the $\delta$-function shown above, defining
\be
\Sigma(q)=\int_{-\infty}^{+\infty}\! d\omega\, {\rm Re}\,\sigma_{yy}(q,\omega).
\ee
(In calculating this, we neglect spectral density at high frequencies that would result from real
transitions between Landau levels.) We discuss the non-uniqueness of the $q^2$ coefficient in this
procedure at the end.

There are also deeper reasons to be skeptical of the formula (\ref{mhr_pred2}). The result in eq.\ 
(\ref{hallviscrel}) holds when the bulk of the system is rotation invariant; that symmetry is broken at 
an edge. Further, the proof of the relation with Hall conductivity uses Galilean invariance for
{\em both} the $x$ and $y$ directions. The edge breaks translation invariance, and invariance under 
a Galilean boost, in the direction normal to the edge, though not necessarily along the edge. In the 
arguments of Ref.\ \cite{mhr}, the bulk effects that hold under the symmetry are assumed to persist up to 
the edge; however, due to the breaking of various symmetries at a {\em generic} edge, it is not clear 
that the assumptions will hold in full generality. For a more formal treatment of these issues, 
the effective theory at the edge should be constructed as the most general one consistent with the 
symmetries. Because conservation of particle number is not broken at an edge, the bulk Hall conductivity 
implies the existence of gapless edge degrees of freedom, and of a coefficient $\nu$ as the chiral anomaly 
in the effective theory for the edge. This $\nu$ is the same $\nu$ as in the bulk Hall conductivity, 
which is described by an ordinary Chern-Simons term in the bulk with that coefficient \cite{ch,wen}. In 
the case of the Hall viscosity, the expression (\ref{hallviscrel}) and the relation with the shift, can 
be obtained \cite{rg,hoyos} from the so-called first Wen-Zee term \cite{wz92} in the bulk effective action.
That term resembles the Chern-Simons term in that it is not locally covariant under some types of gauge
transformations, specifically internal rotations (which are related to the physical rotation symmetry)
\cite{br1}. However (and as in the analogous case of the part of the Lorentz group that is broken by an
edge in a relativistic system), rotation symmetry is lost at an edge, so that the analogy with the chiral 
anomaly does not go through \cite{br2}. Indeed, it has recently been shown explicitly \cite{gja}, at least 
in the absence of Galilean invariance, that a local term can be included in the effective action at an edge 
that restores the internal-rotation gauge invariance, without coupling to the gapless degrees of freedom 
on the edge. While there may be more work to be done on the edge theory with Galilean invariance,
so far as is known there is no indication that the theory for the gapless degrees of freedom at a
generic edge will contain the coefficient $\bar{s}$. But we emphasize that these remarks concern a
{\em generic} edge; it is also possible that the effect as claimed does arise in somewhat more special
circumstances, including some in which the spectral density is a $\delta$-function through order $q^2$,
as we will discuss briefly at the end.

In this Comment we will compare the prediction in eq.\ (\ref{mhr_pred2}) with a direct calculation in 
a simple model.
Our main calculation is for non-interacting electrons filling the lowest Landau level, in the presence of 
a confining edge potential; this system has $\nu=1$ and ${\cal S}=1$, and is 
covariant under Galilean transformations along the edge. We will show that $\Sigma(q)$
at order $q^2$ depends on details of the edge potential, and thus is not universal. 
We find
\be
\Sigma(q) = \half v_F\left(1-\frac{1}{2} q^2 +\frac{3}{2}\varepsilon q^2\right)
\ee
through order $q^2$ and $\varepsilon$. Here $\varepsilon=V''(0)\ell_B^2/(2\omega_c)=mV''(0)/2$ is a 
perturbation due to curvature in the edge potential $V(x)$, $\omega_c=B/m$ is the cyclotron frequency, and 
we set $B$ (and the magnetic length $\ell_B$) to $1$. For $\varepsilon=0$ the confining potential is 
linear. In this case, we extend the calculation to the case in which the $\nu$ lowest Landau levels 
are filled, and find $\Sigma(q)=\half\nu v_F(1-\frac{\cal S}{2}q^2)$ plus order $q^4$, in partial 
agreement with the prediction. Finally, we discuss when such a result may hold more generally.

\section{Calculation for non-interacting electrons}
\subsection{Linear response theory}

The conductivity is the linear response of the current density to a position- and time-dependent 
electric field; we represent the electric field with a scalar potential, which couples to the number 
density of the particles. In Fourier space for a two-dimensional system, the $yy$-component of 
the conductivity tensor for the edge at wavevector $q_y=q$ along the edge (along the $y$ direction) 
and frequency $\omega$ is given by
\be
\sigma_{yy}(q,\omega)=-\frac{1}{q}\int_0^\infty dt\,\int dy\, e^{i\omega^+t-iqy}\langle 
[j_y(y,t),\rho(0,0)]\rangle,
\ee
where the current operator $j_y(y,t)$ and number density operator $\rho(y,t)$ are in the Heisenberg 
picture, and have each been integrated over the $x$ coordinate, so
that the conductivity is essentially one dimensional. The average is taken in the ground state of
the time-independent system with no electric field. The real part of $\sigma_{yy}$ is the spectral 
density, and receives contributions from real transitions (caused by the electric field) to 
excited states. There is a large bulk contribution to this conductivity that drops out of the real part 
at low frequency, where only transitions among the low-lying edge states are involved.

\subsection{Model}

The Hamiltonian for a single charged particle in the presence of a magnetic field and a time-independent 
``confining'' potential $V(x)$ can be written in first quantization. For the vector potential $\bf A$ 
representing the magnetic field $B=\partial_xA_y-\partial_yA_x$, we choose Landau gauge ${\bf A}=(0,Bx)$. 
Our confining potential preserves translation-invariance in the $y$ direction, so energy eigenstates are of 
the form $e^{iky}u_k(x)$, and we can write the reduced Hamiltonian for each $k$ as
\be
{\cal H}_k=-\frac{\partial_x^2}{2m} + \frac{1}{2m} (k-x)^2+V(x),
\ee
where we have again set $\hbar$ and $B=1$ (these imply that the magnetic length $\ell_B=\sqrt{\hbar/B}=1$). 
We can impose a periodic boundary condition on $y$ with period
$L$ if desired, in which case the values of $k$ are $2\pi p/L$, with $p$ an integer.

We assume that $V(x)$ is increasing with $x$, and locate the edge close to $x=0$, with particles occupying 
only the lowest Landau level ($n=0$) and only in the region to the left of the edge, and send the left 
edge off to $x=-\infty$, so it is dropped. For calculations of the conductivity response at the edge, 
it will be sufficient to use a Taylor series expansion for the potential in that region. Hence we expand
\be
V(x) = V(0) + V'(0)x+\sum_{r\geq 2}\frac{1}{r!}V^{(r)}(0)x^r.
\ee
The constant $V(0)$ can be dropped, and we set ${\cal H}_k={\cal H}_k^{(0)}+{\cal H}_k'$,
\bea
{\cal H}_k^{(0)}&=&-\frac{\partial_x^2}{2m}+\frac{1}{2m} (k-x)^2 +V'(0)x,\\
{\cal H}_k'&=&\sum_{r\geq 2}\frac{1}{r!}V^{(r)}(0)x^r.
\eea
We write $E_{nk}$ and $u_{nk}(x)$ for the energy eigenvalues and normalized real eigenfunctions of 
${\cal H}_k$.

${\cal H}_k^{(0)}$ can be written
\be
{\cal H}_k^{(0)}=-\frac{\partial_x^2}{2m} + \frac{1}{2m} \left[\left(k-mV'(0)\right)-x\right]^2 + V'(0)k - 
\frac{m}{2} V'(0)^2.
\ee
Its energy eigenvalues and normalized eigenfunctions are 
\bea
E^{(0)}_{n,k}&=&\omega_c(n+\frac{1}{2}) + V'(0)k - \half m V'(0)^2,\\
u_{n,k}^{(0)}&=&\frac{1}{\sqrt{2^n n! \sqrt{\pi}}} H_n(x-X_k) e^{-(x-X_k)^2/2},
\eea
where $H_n(x)$ are Hermite polynomials, and we define $X_k = k-mV'(0)$. We see that the Landau levels, 
labeled by $n=0$, $1$, $2$, \ldots, are tilted, so their group velocity (in the positive $y$ direction) 
is $v_k^{(0)}=V'(0)$, independent of $k$. Later, part of ${\cal H}_k'$ will be treated as a perturbation 
to this; this will modify the linear relation of $E_{n,k}$ with $k$ for each $n$. We emphasize that 
${\cal H}_k^{(0)}$ should not be taken too literally as a model for the whole system (which would mean 
that infinitely many Landau levels are occupied further to the left in $x$, if a value of the chemical 
potential is given); rather it will be useful for the conductivity in the edge region.

We note that the Hamiltonian ${\cal H}_k$ transforms covariantly under Galilean boosts along the $y$ 
direction. The effect is to add a constant electric field in the $x$-direction; with a choice of gauge, 
this just changes the slope of the linear potential term. This corresponds to adding the velocity by 
which the system was boosted to the velocities of excitations, as expected.

\subsection{Spectral weight}

The $\omega$ integral of the real part of the edge conductivity, or spectral weight,
can be reduced to
\be
\Sigma(q)=\frac{1}{2q}\int_{k_F-q/2}^{k_F+q/2}dk\,j_{k-q/2,k+q/2}\rho_{k-q/2,k+q/2}
\ee
where the matrix elements of the current $j=j_y$ and density are
\bea
j_{k,k+q}&=&\frac{1}{m}\int dx\,(k+\half q-x)u_{0k}(x)u_{0k+q}(x),\\
\rho_{k,k+q}&=&\int dx\,u_{0k}(x)u_{0k+q}(x).
\eea
Here, the terms involving the levels with $n>0$ have been omitted. We placed the Fermi wavevector 
at $k=k_F$, that is all states with $k<k_F$ have been filled to form the ground state, and we took 
the $L\to\infty$ limit for the $y$-direction. Thus, only transitions involving creation of an electron 
and a hole in the lowest Landau level have been retained. 

\subsection{Linear potential case}

For the linear potential as in ${\cal H}_k^{(0)}$, it is straightforward to calculate $\Sigma(q)$. Using 
the zeroth order $u_{0k}^{(0)}$ in place of $u_{0k}$, one finds (with $v_F^{(0)}=v_k^{(0)}=V'(0)$)
\be
j_{k,k+q}^{(0)}=v_F^{(0)}\rho_{k,k+q}^{(0)}=v_F^{(0)}e^{- q^2/4},
\ee
and
\be 
\Sigma(q) = \half v_F^{(0)} e^{-q^2/2}.
\ee
These are independent of $k_F$, and (given the omission of the $n>0$ levels from the spectrum) 
are exact for all $q$. The Gaussian dependence on $q$ comes from the overlap of Gaussians with their 
centers separated by $q$. We also note that in the present case, the spectral density is exactly 
a $\delta$-function at $\omega=qv_F^{(0)}$, with coefficient $\Sigma(q)$.

\subsection{Case of $\nu$ filled Landau levels}

It is not difficult to generalize the preceding calculation to non-interacting electrons with $\nu$ Landau 
levels ($\nu$ is an integer) filled in the bulk and a linear potential in the edge region. We can ask if 
the dependence on $\nu$ and $\cal S$ resembles that in the prediction eq.\ (\ref{mhr_pred1}).

For low frequency and wavevector, the transitions that contribute to the spectral weight $\Sigma(q)$
are those in which the electron stays within the same Landau level. The matrix elements of the density 
and current for transitions within the $n$th level, instead of the zeroth, again obey 
$j_{k,k+q}=v_F^{(0)}\rho_{k,k+q}$, with $v_F^{(0)}$ independent of $n$. The matrix elements of
the density within the $n$th level are found to be (similar to Ref.\ \cite{haldane})
\bea
\rho_{k,k+q}&=&e^{-q^2/4}\sum_{r=0}^n\frac{1}{r!}{n \choose r}(-\half q^2)^r\\
&=&e^{-q^2/4} L_n(\half q^2)\\
&=&1-\half(n+\half)q^2+O(q^4)
\eea
independent of $k$; $L_n(x)$ are the Laguerre polynomials [the generating function
\be
e^{2xt-t^2}=\sum_{n=0}^\infty H_n(x)\frac{t^n}{n!}
\ee
is useful in carrying out the exact calculation]. Then for
the spectral weight we find
\bea
\Sigma(q)&=&\half v_F^{(0)}e^{-q^2/2}\sum_{n=0}^{\nu-1}[L_n(\half q^2)]^2\\
&=&\half\nu v_F^{(0)}\left(1-\bar{s}q^2\right)+O(q^4),
\label{our_res}
\eea
where we used the fact that the mean orbital spin per particle $\bar{s}$ for the state in which $\nu$
levels are filled is given by
\be
\nu\bar{s}=\sum_{n=0}^{\nu-1}(n+\half),
\ee
so $\bar{s}=\half\nu$ \cite{wz92}.
Thus the result for a linear potential does have the form of the prediction, in that $q^2$ is multiplied
by ${\cal S}=2\bar{s}$, but the numerical coefficient $1/4$ should be replaced by $-1/2$.

\subsection{Non-linear potential}

For the full non-linear potential $V(x)$, the coefficients in the Taylor series $V^{(r)}(0)$, $r\geq 2$, 
are independent perturbation parameters. We will consider only the first one $V''(0)$, thus adding a 
quadratic perturbing potential ${\cal H}_k^{(1)}=\half V''(0)x^2$ to ${\cal H}_k^{(0)}$, and for simplicity 
consider only $\nu=1$.

Using standard first-order perturbation theory, we find for the change in the energy eigenvalues and 
normalized eigenfunctions (for $n=0$ only) to first order in $\varepsilon=mV''(0)/2$
\bea
\Delta E_{0k}&=&\half V''(0)(X_k^2+\half),\\
\Delta u_{0k}(x)&=&-\sqrt{2}\varepsilon X_ku_{1k}^{(0)}(x)-\frac{1}{2\sqrt{2}}\varepsilon u_{2k}^{(0)}(x)
\eea
(the change in normalization coefficient for the zeroth order part of $u_{0k}$ is second order in 
$\varepsilon$ and can be omitted). $\Delta E_{0k}$ is quadratic in $k$, so produces curvature in the 
dispersion relation, as expected. We now set $k_F=0$; the velocity at $k=0$ including the perturbation 
to first order is then
\be
v_F=V'(0)(1-2\varepsilon).
\ee
In $\Delta u_{0k}$, the first term represents a shift to the left
of $u_{0k}^{(0)}$ that is proportional to $X_k$, and so it decreases the spacing of the centers of the 
$u_{0k}$. At the same time, the second term decreases the width of each function. These effects are to 
be expected from an edge potential that increases faster than linearly as $x$ increases \cite{halperin}.

Using the eigenfunctions to order $\varepsilon$ to calculate the matrix elements, we find
\bea
\rho_{k,k+q}&=&\left(1+\frac{3}{4}\varepsilon q^2\right) e^{-q^2/4},\\
j_{k,k+q}&=&v_F^{(0)}\rho_{k,k+q}+\frac{2\varepsilon}{m}X_{k+q/2}e^{-q^2/4}.
\eea
In $\rho_{k,k+q}$, the two terms in $\Delta u_{0k}$ have opposite effects of order $\varepsilon q^2
e^{-q^2/4}$, but the decrease in the spacing of the centers of the wavefunctions, which increases 
their overlaps, dominates. In $j_{k,k+q}-v_F^{(0)}\rho_{k,k+q}$, the effect comes entirely from the 
decrease in the spacing of the centers. Part of the effect on $j_{k,k+q}$ is to replace $v_F^{(0)}$ 
by $v_F$ (to first order in $\varepsilon$) times $\rho_{k,k+q}$.

Finally, we find that in $\Sigma(q)$, $j_{k,k+q}=v_F\rho_{k,k+q}$ can be used through order $\varepsilon$, 
and we obtain the low-frequency spectral weight
\be
\Sigma(q)=\half v_F \left(1+\frac{3}{2}\varepsilon q^2\right)e^{-q^2/2}
\ee
plus order $\varepsilon^2$. If desired this may be expanded to order $q^2$ to obtain the result quoted 
above.

\section{Discussion}

Here we first ask whether, through order $q^2$, it is sufficient to study the spectral weight $\Sigma(q)$ 
rather than the full spectral density $\sigma'(q,\omega)\equiv {\rm Re}\,\sigma_{yy}(q,\omega)$; again, 
we consider only $\nu=1$. First,
if the energy dispersion $E_{0k}$ is analytic and its Taylor expansion about $k=k_F$ contains a 
quadratic term [i.e.\ $(k-k_F)^2$] with positive coefficient, as well as the linear $v_F(k-k_F)$, then 
for small $q$ the spectral density at positive $q$ is non-zero for $\omega$ between $E_{0k_F}-E_{0k_F-q}$ 
and $E_{0k_F+q}-E_{0k_F}$, a range of order $q^2$, and is zero outside this range. (Here, as throughout 
this Comment, we once again ignore the spectral density outside the low-frequency region.)

We now examine moments $\int d\omega\,\omega^p\sigma'(q,\omega)$ of the spectral density ($p=0$, $1$, 
\ldots), and define
\be
\overline{\omega}=\frac{\int d\omega\,\omega\sigma'(q,\omega)}{\int d\omega\,\sigma'(q,\omega)},
\ee
which is of course of order $q$ as $q\to0$.
We can reconstruct $\sigma'(q,\omega)$ from its moments. But by expanding in powers of 
$\omega-\overline{\omega}$, we have
\bea
\overline{\omega}^{-p}\int d\omega\,\omega^p\sigma'(q,\omega)&=&\int d\omega\,\sigma'(q,\omega)\nonumber\\
&&{}+{p\choose 2}\overline{\omega}^{-2}
\int d\omega\,(\omega-\overline{\omega})^2\sigma'(q,\omega)\nonumber\\
&&{}+\ldots
\eea
(where \ldots denotes terms with higher moments of $\omega-\overline{\omega}$), and the second term on 
the right is upper bounded by order $q^2$ times the first term as $q\to0$, and the first term is 
$\Sigma(q)$, of order $1$. Because of positivity, there can be no cancelations in the second term, and 
the result will usually be of order $q^2$. Hence different estimates for the coefficient of a 
$\delta$-function, such as use of different moments, can give different results at order $q^2$; it is not 
clear what the prediction of Ref.\ \cite{mhr} is supposed to mean in this (generic) situation. However, 
in particular cases, one of these estimates could agree with the prediction from Ref.\ \cite{mhr}.

If we now consider including interactions, then the spectral density will usually be broadened, even when 
the confining potential is linear in the edge region. For example, for the case of $\nu$ filled Landau 
levels, a first approximation could use the Hartree-Fock method, and in general this produces curvature (and
different velocities for the different Landau levels) in the energy dispersion, even when the background 
potential is linear (see Ref.\ \cite{glazman} for further discussion). 

However, it might be that the same form, eq.\ (\ref{our_res}), holds (through order $q^2$) for other 
states in the presence of the linear potential, for some Hamiltonians. In particular, there are 
certain ``special'' Hamiltonians for each of which some model wavefunction---such as Laughlin's---is 
the exact ground state, there is believed to be a gap in the bulk energy spectrum, and the edge 
excitations all move with the same velocity or angular velocity \cite{read09b}. In these models, the 
spectral density at small $q$ and $\omega$ is again a $\delta$-function, even though there are 
interactions, and in view of the nice forms of the wavefunctions of the edge excitations, which are 
known exactly, we might expect to find the same form (\ref{our_res}) for the $q^2$ part of the 
conductivity, with $\bar{s}={\cal S}/2$ related to the shift of the state.

\section{Conclusion}

We have shown that, contrary to the claim in Ref.\ \cite{mhr}, the $q^2$ term in the real part of the 
longitudinal conductivity at the edge of a quantum Hall system is
not universal. Indeed, the form for the real part (the spectral density) with a $\delta$-function at 
$\omega=qv_F$ cannot generally be employed at order $q^2$, and no procedure for mapping a generic result 
onto the $\delta$-function form was given by the authors of Ref.\ \cite{mhr}. For the integral of the 
spectral density, the coefficient of $q^2$ depends on details of the confining potential. 
For the special case in which the potential is linear, the results in some model states are similar 
to the prediction, though with a different coefficient times the shift. Thus the predicted form 
(\ref{mhr_pred2}) may hold [after it is corrected as in eq.\ (\ref{our_res})] under 
conditions that involve (i) a potential that is linear near the edge and (ii) particular forms for 
the interactions (if present) such that the spectral density can be represented by a $\delta$-function 
through order $q^2$, but not under more general conditions.

\acknowledgements

We are grateful to S. Moroz, L. Radzihovsky, L.I. Glazman, and B. Bradlyn for their comments on an 
earlier version of this Comment. This research was supported by NSF grant No.\ DMR-1408916.


\begin{references}

\bibitem{asz} J.E. Avron, R. Seiler, and P.G. Zograf, Phys. Rev.
Lett. {\bf 75}, 697 (1995).

\bibitem{read09} N. Read, Phys. Rev. B {\bf 79}, 045308 (2009).

\bibitem{rr11} N. Read and E. Rezayi, Phys. Rev. B {\bf 84}, 085316 (2011).

\bibitem{wz92} X.-G. Wen and A. Zee,
Phys. Rev. Lett. {\bf 69}, 953 (1992); {\it ibid.} {\bf 69}, 3000 (1992) (E).

\bibitem{hoyos} C. Hoyos and D.T. Son, Phys. Rev. Lett. {\bf 108}, 066805 (2012).

\bibitem{bgr} B. Bradlyn, M. Goldstein, and N. Read, Phys. Rev. B {\bf 86}, 245309 (2012).

\bibitem{mhr} S. Moroz, C. Hoyos, and L. Radzihovsky, Phys. Rev. B {\bf 91}, 195409 (2015).

\bibitem{wen} X.-G. Wen, Int. J. Mod. Phys. B {\bf 06}, 1711 (1992).

\bibitem{ch} C.G. Callan and J.A. Harvey, Nucl. Phys. B {\bf 250}, 427 (1985).

\bibitem{halperin} B.I. Halperin, Phys. Rev. B {\bf 25}, 2185 (1982).

\bibitem{rg} N. Read and W. Goldberger, (2009), unpublished.

\bibitem{br1} B. Bradlyn and N. Read, Phys. Rev. B {\bf 91}, 125303 (2015). 

\bibitem{br2} B. Bradlyn and N. Read, Phys. Rev. B {\bf 91}, 165306 (2015).

\bibitem{gja} A. Gromov, K. Jensen, and A.G. Abanov, Phys. Rev. Lett. {\bf 116}, 126802 (2016).

\bibitem{haldane} F.D.M. Haldane, in {\it The Quantum Hall Effect}, 2nd Ed., edited by R.E. Prange and 
S.M. Girvin (Springer-Verlag, New York, NY, 1990).

\bibitem{glazman} A. Imambekov, T.L. Schmidt, and L.I. Glazman, Rev. Mod. Phys. {\bf 84}, 1253 (2012).

\bibitem{read09b} See e.g.\ N. Read, Phys. Rev. B {\bf 79}, 245304 (2009).

\end{references}
\end{document}